\def\lae{\mathrel{<\kern-1.0em\lower0.9ex\hbox{$\sim$}}}
\def\gae{\mathrel{>\kern-1.0em\lower0.9ex\hbox{$\sim$}}}

\newcommand{\be}{\begin{equation}}
\newcommand{\ee}{\end{equation}}

\documentclass[apj]{emulateapj}

\usepackage{apjfonts,lscape}

\slugcomment{Accepted for publication in ApJ Letters}
\lefthead{JORD\'AN ET AL.}
\righthead{LMXBs and GCs in Cen~A}

\begin{document}
 
\title{Low-Mass X-ray Binaries and Globular Clusters in Centaurus~A
       \altaffilmark{1}}
\author{
Andr\'es   Jord\'an\altaffilmark{2,3,4},
Gregory R. Sivakoff\altaffilmark{5},
Dean E.    McLaughlin\altaffilmark{6},
John P.    Blakeslee\altaffilmark{7},
Daniel A.  Evans\altaffilmark{2},
Ralph P.   Kraft\altaffilmark{2},
Martin J.  Hardcastle\altaffilmark{8},
Eric W.    Peng\altaffilmark{9},
Patrick    C\^ot\'e\altaffilmark{9},
Judith H.  Croston\altaffilmark{8},
Adrienne   M. Juett\altaffilmark{10},
Dante      Minniti\altaffilmark{4},
Somak      Raychaudhury\altaffilmark{2,11},
Craig L.   Sarazin\altaffilmark{10},
Diana M.   Worrall\altaffilmark{12,2},
William E. Harris\altaffilmark{13},
Kristin A. Woodley\altaffilmark{13},
Mark       Birkinshaw\altaffilmark{12,2},
Nicola J.  Brassington\altaffilmark{2},
William R. Forman\altaffilmark{2},
Christine  Jones\altaffilmark{2},
Stephen S. Murray\altaffilmark{2}
}

\begin{abstract}
We present results of {\it Hubble Space Telescope} and 
{\it Chandra X-ray Observatory} 
observations of globular clusters (GCs)
and low-mass X-ray binaries (LMXBs) in the central regions of Centaurus~A.
Out of 440 GC candidates we find that 41 host X-ray point sources 
that are most likely LMXBs. 
We fit King models to our GC candidates in order to measure their 
structural parameters.
We find that GCs that host LMXBs are denser and more compact, and have higher encounter 
rates and concentrations than the GC population as a whole. 
We show that the higher
concentrations and masses are a consequence of the dependence of LMXB 
incidence on central density and size plus the general trend for denser
GCs to have higher masses and concentrations.
We conclude that neither concentration nor mass are fundamental variables
in determining the presence of LMXBs in GCs, and that 
the more fundamental parameters 
relate to central density and size.
\end{abstract}


\altaffiltext{1}{Based on observations with the NASA/ESA
{\it Hubble Space Telescope} obtained at STScI,which is operated
by AURA, Inc., under NASA contract NAS 5-26555}
\altaffiltext{2}{Harvard-Smithsonian Center for Astrophysics, 
Cambridge, MA 02138; \{ajordan, devans, rkraft, wforman, cjones, ssm\}@cfa.harvard.edu}
\altaffiltext{3}{Clay fellow}
\altaffiltext{4}{Departamento de Astronom\'{\i}a y Astrof\'{\i}sica, 
P.\ Universidad Cat\'olica de Chile, Casilla 306, Santiago 22, Chile; dante@astro.puc.cl}
\altaffiltext{5}{Department of Astronomy, The Ohio State University, 
Columbus, OH 43210-1173; 
sivakoff@astronomy.ohio-state.edu}
\altaffiltext{6}{School of Physical and Geographical Sciences, Astrophysics Group, 
Keele University, Keele, ST5 5BG, UK; dem@astro.keele.ac.uk } 
\altaffiltext{7}{Department of Physics and Astronomy,
Washington State University, Pullman, WA 99163-2814; jblakes@wsu.edu}
\altaffiltext{8}{School of Physics, Astronomy, and Mathematics, 
University of Hertfordshire, Hatfield AL10 9AB, UK; \{m.j.hardcastle, j.h.croston\}@herts.ac.uk}
\altaffiltext{9}{Herzberg Institute of Astrophysics, Victoria, 
BC V9E 2E7, Canada; \{eric.peng, patrick.cote\}@nrc-cnrc.gc.ca}
\altaffiltext{10}{Department of Astronomy, University of Virginia, P. O. Box 400325, 
Charlottesville, VA 22904-4325; \{juett, sarazin\}@virginia.edu}
\altaffiltext{11}{School of Physics and Astronomy, University of Birmingham, 
Birmingham B15 2TT, UK; somak@star.sr.bham.ac.uk}
\altaffiltext{12}{ Department of Physics, University of Bristol, Bristol BS8 ITL, UK; 
\{diana.worrall, mark.birkinshaw\}@bristol.ac.uk}
\altaffiltext{13}{Department of Physics and Astronomy, McMaster University, 
Hamilton, ON L8S 4M1, Canada; \{harris, woodleyka\}@physics.mcmaster.edu}

\section{Introduction}
\label{sec:intro}

Since the discovery of X-ray sources associated with globular clusters (GCs) 
in the Milky Way  (MW; e.g., Giacconi et~al.\ 1974) 
it has been known that low-mass X-ray binaries (LMXBs) 
are formed more efficiently in GCs by a factor of ~100 compared to the field
(Katz 1975; Clark 1975). 
This is thought to be a direct consequence of the high central densities in GCs.  
At high density, dynamical formation mechanisms 
are greatly enhanced relative to the field (Clark 1975; Fabian et~al.\ 1975; Hills 1976).

While the dynamical origin of most GC LMXBs was proposed three
decades ago, subsequent observational studies 
of the core properties of GCs that host LMXBs have
been largely restricted to the MW. 
The {\it Chandra X-ray Observatory} has made it possible to study the    
LMXB populations of nearby
($\lesssim 30$ Mpc) galaxies 
(e.g., Fabbiano 2006 and references therein), but the
core properties of the GCs, which are thought to be the most relevant
in the dynamical processes that create LMXB progenitors, are not
robustly determined even with {\it HST} for most of those galaxies.

In the MW, Bellazzini et~al.\ (1995) showed that 
bright LMXBs appear in denser and more metal-rich GCs,  while Pooley et~al.\
(2003) and Heinke et~al.\ (2003) 
showed that the number of GC X-ray sources correlates with $\Gamma$, a basic
indicator of dynamical encounter rates defined as 
$\Gamma\equiv \rho_0^{1.5} r_c^2$ (e.g., Verbunt et~al.\ 2007), 
where $\rho_0$ is the central mass density and 
$r_c$ is the core radius.
Bregman et~al.\ (2006) showed additionally that bright GC LMXBs in the MW
appear preferentially in GCs with smaller core and half-light radii and 
shorter half-mass relaxation times.
Sivakoff et~al.\ (2007; S07) used 11 early-type galaxies to show that 
encounter rates are a good predictor for the appearance of LMXBs in GCs,
deriving $\Gamma$ from measurements at the half-light 
radii. This is consistent with earlier results in M87 that used more 
uncertain derived core properties (Jord\'an et~al.\ 2004; J04). 

To make further progress in understanding the dynamical processes that 
create LMXB progenitors in the cores of GCs it is necessary to directly
observe LMXBs and the core structure of the GCs that host them 
for as large a sample as possible. 
In this work, we use {\it HST/ACS} 
and {\it Chandra} observations of
the central regions of 
Centaurus~A  (or NGC~5128; hereafter Cen~A)
in order to extend such studies beyond the Local Group.
At the distance of Cen~A, $D\approx 3.7$ Mpc 
(average of 5 distance indicators, see \S6 in Ferrarese et al 2007),
it is possible to study the central properties of GCs
with {\it HST}, an ability we exploit in what follows in order to 
study which structural properties of GCs are most relevant
in determining the presence of LMXBs in GCs.

\section{Observations}
\label{sec:data}

{\it Optical Catalog.} We have made use of 21 fields in Cen~A
observed with the F606W filter using the Wide Field Channel 
mode of the {\it Advanced Camera for Surveys} (ACS) on board the {\it Hubble
Space telescope} (HST). Nine of these fields were observed as part
of program GO-10597 (PI: A. Jord\'an), which was designed to be
combined with previous observations from program GO-10260 (PI:
W.E. Harris) in order to cover a large area in the inner
$r\lesssim 8$~kpc plus some fields at larger radii.

Each field of GO-10597 consists of four 525-sec exposures plus a
single 58-sec exposure, while the data of GO-10260 consist of
three exposures of 790 sec. Full observational details of
GO-10260 are given in Harris et~al.\ (2006), but we note that
for this study we have re-reduced the GO-10260 data in parallel
with our reductions of the new GO-10597 data to construct an
independent and homogeneous catalog of GCs. A detailed account of 
the data reduction procedures and GC catalog construction will 
be given elsewhere so only a brief summary follows here.

All data were drizzled using the Apsis package (Blakeslee et~al.\ 2003) 
onto frames with a pixel scale of 0\farcs05. Object
detection and photometry were performed using
SExtractor (Bertin \& Arnouts 1996). Our detection threshold
corresponds to $m_{\rm F606W}\sim 22$ AB mag and
is such that we are not subject to incompleteness effects. We
used photometric zeropoints and extinction coefficients from
Sirianni et al.\ (2005) and de-reddened the photometry using
$E(B-V)=0.115$ (Schlegel et~al.\ 1998). We dealt with
foreground-star contamination by eliminating all objects
that were consistent with the point-spread function (PSF). We
matched our object catalog with the full photometric catalog (56,674 objects) 
of Peng et~al.\ (2004), obtaining $VI$ photometry
for all objects for which a match was found. The same catalog was
used to place our objects on a consistent astrometric frame. For
objects which did not have a match in Peng et al., we further
matched to deep $VI$ photometry of the central regions of Cen~A
obtained with the VLT (Minniti et~al.\ 2004).

An azimuthally averaged surface-brightness profile was obtained
for each object  using the
ELLIPSE task in IRAF. We then fit PSF-convolved King (1966)
models to each object following the procedures applied
by McLaughlin et~al.\ (2007) to the GO-10260 data 
(see also Barmby et~al.\ 2007). 
The best-fit models were used to infer various global and
core parameters for every GC. 
\footnote{At the distance of Cen~A, a typical GC core radius of $r_c \sim 1$~pc
corresponds to 0\farcs056, comparable to the half-width at
half-maximum of the PSF.  Holland, C\^ot\'e, \& Hesser
(1999) and Harris et~al.\ (2002) have shown that estimates of
$r_c$ from PSF-convolved King-model fitting in this situation are
unbiased in general, and Barmby et~al.\ (2007) have shown
that the fits of McLaughlin et~al.\ (2007) 
to the GO-10260 GCs in Cen~A, which include $\sim 25\%$
of the GCs in our sample, give rise to core-parameter correlations
that are consistent with those found for MW GCs.} 

To infer mass-based cluster quantities 
we used $F606W$ mass-to-light ratios
as a function of $(V-I)$ color derived from the
models of Bruzual \& Charlot (2003) assuming
an age of 13 Gyr for all objects\footnote{For objects without
measured $(V-I)$ we used $(M/L)_{\rm F606W} = 2.2$ in solar
units.}. We culled our catalog by rejecting objects with
$(V-I) < 0.6$, $(V-I) > 1.55$ or concentration $c > 2.5$. 
We further eliminated objects whose surface-brightness profiles
spanned less than 0.5 mag~arcsec$^{-2}$ or that were
visually deemed to be background objects.
The final catalog consists of 440 objects, 407 of
which have VI photometry available. 
We estimate that $\sim\!3\%$ of these
objects may be contaminating background galaxies (based on an
analysis of 21 control fields observed in the same filter to
similar depths, and analyzed in the same fashion as our Cen~A
program fields).

{\it X-ray Catalog.} Prior to 2007, {\it Chandra} 
observed Cen A with the ACIS detectors four times
(Observations
\dataset[ADS/Sa.CXO#obs/00316]{316},
\dataset[ADS/Sa.CXO#obs/00962]{962},
\dataset[ADS/Sa.CXO#obs/02978]{2978},
and
\dataset[ADS/Sa.CXO#obs/03965]{3965}). In 2007,
{\it Chandra} performed
six deep observations ($\sim 100 {\rm \, ks}$ each) of Cen~A 
as part of the
Cen~A Very Large Project (CenA-VLP; PI: R. Kraft). 
In this analysis, we report initial results 
from an X-ray point source list constructed 
using the first four CenA-VLP observations (Observations
\dataset[ADS/Sa.CXO#obs/07797]{7797},
\dataset[ADS/Sa.CXO#obs/07798]{7798},
\dataset[ADS/Sa.CXO#obs/07799]{7799}, and
\dataset[ADS/Sa.CXO#obs/07800]{7800}) plus the previous
observations listed above. 

The reduction of the {\it Chandra} data prior to 2007
is described in Woodley et~al.\ (2007). 
The reduction of the CenA-VLP observations (a total of 373,353 s) will be 
presented in detail	
in forthcoming papers; here, we only consider a few aspects relevant to 
point source detection
(see Hardcastle et~al.\ 2007 for an image of the X-ray data).
Sources were detected 
in the 0.5--7~keV X-ray image from each observation using {\sc wavdetect} 
with wavelet scales spaced by a factor of $\sqrt{2}$ and ranging	
from 1 to 32 pixels, with a source detection threshold of $10^{-7}$. 
Sources that were not associated with the X-ray jet or lobes in
Cen~A were used to register the relative astrometry between the observations. 
The majority of the detected sources are X-ray binaries in Cen~A, 
and in particular the ones
associated with GCs are expected to be LMXBs. In what follows we will
use the term LMXBs for X-ray sources associated with GCs.

{\it Source Matching.} 
After determining the astrometric offsets required to bring the astrometry of 
our list of X-ray sources into the same system as the GCs, we carried
out source matching with a matching radius of $1\farcs6$, obtaining 41 matches. 
With our chosen matching radius we expect $\lesssim 1$ false matches.
The {\it rms} difference between the X-ray and optical positions
for matched sources was $\lesssim 0\farcs5$ in both right ascension and declination.

\section{Results}
\label{sec:res}

\begin{figure*}
\epsscale{1}
\plotone{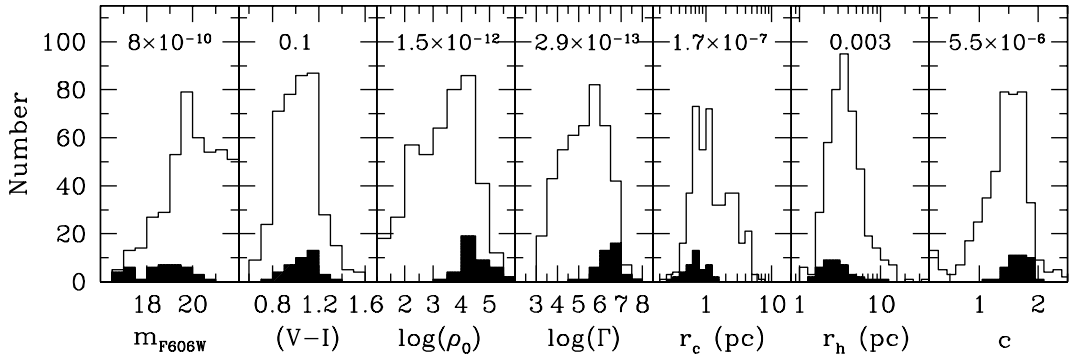}
\caption{
This figure shows histograms of total apparent F606W AB magnitude ($m_{\rm F606W}$), 
de-reddened $(V-I)$ color, 
logarithm of the central mass density ($\rho_0$, in units of $M_\odot$ pc$^{-3}$)
and encounter rate ($\Gamma\equiv \rho_0^{1.5}r_c^2$), King (1966) core radii ($r_c$),
half-light ($r_h$) radii and concentration ($c$) for the full sample of 
GCs (unfilled histograms) and for the subsample that coincide with an X-ray source
(filled histograms). The numbers in each of the panels of this figure show the $p$-value
of a two-sample Wilcoxon test between the two 
distributions shown. The $p$-value gives the probability that the 
two distributions have the same mean values.
\label{fig:dyn_g}}
\end{figure*}

\placefigure{\ref{fig:dyn_g}}

Previous studies 
have found that LMXBs appear preferentially
in more luminous (massive) and redder (metal-rich) GCs 
(e.g.\ S07; Kundu et~al.\ 2007; see Fabbiano 2006 
for a recent review and Minniti et~al.\ 2004 and Woodley et~al.\ 2007
for studies in Cen~A). 
Metal-rich GCs are $\approx 3$
times more likely to host a LMXB, although the scatter around this
value for different galaxies is significant.

In the leftmost panel of Figure~\ref{fig:dyn_g} we show the histogram of 
AB magnitude in the F606W band for all GC candidates (unfilled histogram)
and for those that have an associated X-ray point source 
(filled histogram). The next panel shows equivalent histograms in
$(V-I)$ color for sources with $VI$ photometry. 
The numbers in each of the panels of this figure show the $p$-value
of a two-sample Wilcoxon test between the two 
distributions shown. The $p$-value gives the probability that the 
two distributions have the same mean values.
It is clear
that there is a strong preference for LMXBs to be hosted by more
luminous (massive) GCs. The color distribution of LMXB
hosting GCs is slightly more weighted toward redder GCs,
although the difference between the mean of the two distributions is not significant
for our sample.

In the five
rightmost panels of Figure~\ref{fig:dyn_g} we show
histograms of central mass density ($\rho_0$), 
core encounter rate ($\Gamma$),
King (1966) core ($r_c$) and half-light ($r_h$) radii and 
concentration $c\equiv\log(r_t/r_c)$, where $r_t$ is the model tidal-radius,
for all GCs (unfilled histograms) and those that host an X-ray point
source (filled histograms). In agreement with observations in
elliptical galaxies 
(J04; S07) and the MW (Bregman et~al.\ 2006), we find that the GCs hosting LMXBs have 
smaller half-light radii and higher encounter rates.  
Our new observations allow us to show explicitly for the first time in
an elliptical galaxy that they also have significantly smaller core radii and higher
central densities, as is the case in the MW (Bellazzini et~al.\ 1995; 
Bregman et~al.\ 2006).
Finally, the concentration is
also found to be higher in LMXB-hosting GCs (but see below).

It is well known that many of the structural properties of GCs
in the MW are correlated with one another (e.g., see
Djorgovski \& Meylan 1994; McLaughlin 2000). It is therefore
important to ask which of the parameters in Figure~\ref{fig:dyn_g} might be
fundamental in determining the presence or absence of LMXBs in
GCs, and which parameters might just appear to be important
because they correlate with more physically relevant
properties. Of particular interest is disentangling the
degeneracy between central density (and, thus, the encounter rate
$\Gamma \equiv \rho_0^{1.5} r_c^2$) and total mass, which arises
from the fact that massive GCs are denser on average (McLaughlin
2000; Jord\'an et al. 2005). In principle, other quantities
of relevance to LMXB production---such as, e.g., the
fraction of primordial binaries---might depend on GC
mass but not on either of $\rho_0$ or $\Gamma$ on its own, which
leaves the implications of Figure~\ref{fig:dyn_g} somewhat unclear.

We may empirically probe which GC properties influence the
presence of LMXBs in GCs {\it after taking into account any
dependences on total mass}, by comparing the properties of
clusters that host LMXBs with the properties of clusters that do
not host LMXBs but have the same underlying mass distribution. To
do this, we first estimate the mean dependence of each 
of $Y=\{\rho_0, \Gamma, r_c, r_h, c\}$ on GC mass $M$ for GCs
that do not host an LMXB using  a robust local smoothing of the $M$--$Y$ scatterplot
performed with the Lowess method (Cleveland 1979).
We then use the functions $\hat{Y}(M)$ thus estimated to predict the expected average
of the variable $Y$ when having the same mass distribution as the GCs that host LMXBs
by computing $\langle Y \rangle = N^{-1}\sum_{i=1}^N\hat{Y}(M_i)$, where
$M_i$ are the masses of the $N=41$ GCs that host LMXBs.

\begin{figure*}
\epsscale{1}
\plotone{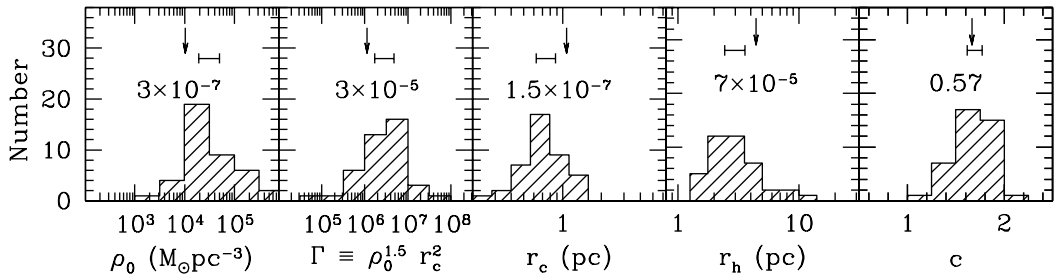}
\caption{
This figure shows histograms 
of the central mass density ($\rho_0$), encounter rate ($\Gamma\equiv \rho_0^{1.5}r_c^2$), 
King (1966) core radii ($r_c$), half-light ($r_h$) radii and concentration ($c$) for 
the full the sample of GCs that are associated with an X-ray point source (X-GCs). The error bar 
in each panel shows a 99\% confidence interval for the median of each distribution and 
the arrow indicates an estimate of the expected value of that median for GCs that do 
not hold LMXBs (nX-GCs) and that have the same undelying mass distribution as the GCs shown in
the histogram.
The number in each panel is the $p$-value of a 
one-sample Wilcoxon test to test the probability that the sample median
is consistent with the value indicated by the arrow.
The ratio of the median values of the quantities that show 
significantly different distributions
(in the sense X/nX) are: 2.1 ($\rho_0$), 1.93 ($\Gamma$), 0.68 ($r_c$) 
and 0.77 ($r_h$).
\label{fig:dyn_M}}
\end{figure*}

We show the result of this exercise in Figure~\ref{fig:dyn_M}, where we show
various distributions of structural quantities for 
GCs that host LMXBs, indicating with an error bar the 99\% confidence interval
for the median of these distributions and with an arrow the expected median 
for GCs that do not host an LMXB and have the same underlying 
mass distribution.
The result of a one-sample Wilcoxon test to estimate the probability that the 
median of the LMXB-hosting sample is consistent with that expected for GCs that do not
host LMXBs is indicated in each panel.

A very interesting result is given by the rightmost panel in 
Figure~\ref{fig:dyn_M}: after the effects of mass are taken into account, the 
average King concentration shows no statistical difference between GCs
that host LMXBs and those that do not. In the MW there is a trend for more massive GCs to 
be more concentrated, roughly following $10^c \propto M^{0.4}$ (McLaughlin 2000).
A consistent trend is also present in our sample.
We have shown that 
once this trend is taken into account {\it concentration per se has no significant 
effect in determining the presence of LMXBs}. 

Figure~\ref{fig:dyn_M} also shows that all of $\rho_0$, $\Gamma$, $r_c$, and
$r_h$ have mean values that are significantly different for GCs
that host LMXBs, {\it even when considering clusters with the
same mass distribution}. Thus, processes driven by high core
densities, and presumably related to high stellar encounter
rates, play a direct role in enhancing the presence of LMXBs in
GCs; the dependences on core properties and $r_h$ in Figure~\ref{fig:dyn_g} are
not purely coincidental side effects of physics or initial
conditions that scale fundamentally with cluster mass only.

Is there a role left for GC mass in determining the presence of an LMXB after
the dependence on $\rho_0$ and $r_c$ (through $\Gamma$) has been taken into account?
We can answer this question by repeating the exercise above but now comparing
the mass distribution of GCs that host LMXBs with those of GCs that 
do not host LMXBs and have the same underlying distribution of $\Gamma$. We find
when doing this that the average mass of GCs that host LMXBs is
statistically indistinguishable to that of GCs that do not 
(Wilcoxon $p$-value of 0.98). 
As there are no remaining differences in the average masses 
after dependencies in structural parameters
have been taken into account through $\Gamma$, we conclude 
that {\it the dependence of LMXB incidence on mass is a consequence
of a more fundamental dependence on central density
and size}. We note also that $c$ shows no significant difference
between GCs that host LMXBs and those that do not when considering the same underlying
$\Gamma$ distributions.

\placefigure{\ref{fig:dyn_M}}

We follow the procedure described in \S4 of S07 to determine
the dependence of the expected number of LMXBs, $\lambda_t$, on GC properties
using a maximum-likelihood method. In particular, we fit for the exponents
in the following assumed forms for $\lambda_t$: 
(1) $\lambda_t \propto M^{\beta} 10^{\delta(V-I)}r_h^{\epsilon}$ and 
(2) $\lambda_t \propto \Gamma^{\eta} 10^{\delta(V-I)}$. 

We find that the values of $\delta$ obtained are not 
significantly different from zero. 
This is consistent with the fact 
that the $(V-I)$
color distribution of GCs that host LMXBs is not significantly different
from the global color distribution in our sample.
When using the functional form (1) we find 
$\beta=1.4\pm0.2$, $\delta=0\pm 0.6$ and $\epsilon=-2.1\pm0.4$, while using form (2)
we find $\delta=-0.3\pm0.6$ and $\eta=0.85\pm0.12$. 
These results are in good agreement
with the results of S07 for Virgo Cluster ellipticals, where the analysis
was restricted to using quantities defined at the half-mass radius.

\section{Conclusions}
\label{sec:conc}

We have shown in this {\it Letter} that GCs in Cen~A that host LMXBs     
have significantly higher central densities, smaller sizes, and    
higher concentrations than the GC population as a whole. The higher    
concentrations are shown to result from the dependence of LMXB     
incidence on central density and size, plus the fact that denser (or     
more massive) GCs are more centrally concentrated on average.

More importantly, we have used our Cen~A sample to compare the 
properties of LMXB-hosting GCs to those that do not but 
have the same underlying mass or $\Gamma$ distribution. We find that GCs that 
have LMXBs have significantly higher central densities, encounter rates
and smaller sizes than GCs of the same mass that do not have LMXBs. 
Interestingly, we find that the mean concentrations are 
indistinguishable once the trend for more massive GCs to have higher
concentrations is removed. As $r_h/r_t$ depends only on $c$ for King models, 
our conclusion agrees with the MW data presented in
Bregman et~al.\ (2006; their Figure~1).

We further show that potential GC mass-dependent processes are not 
fundamental in the formation of GC LMXBs (see also S07; Verbunt et~al.\ 2007). 
It is rather smaller sizes and denser cores, or equivalently higher
values of the core encounter rate $\Gamma$, that are the main drivers.

Finally, our finding that concentration is not important in determining
the presence of LMXBs in GCs validates the use of half-mass densities
and radii to probe the dynamical properties of GCs and their connection
to the presence of LMXBs. This is helpful as core properties are 
uncertain for GCs in galaxies much more distant than Cen~A.
The detailed dependence of LMXB incidence on GC core properties will aid
in further understanding of the interplay of various 
dynamical processes in creating X-ray binaries
(e.g., Banerjee \& Ghosh 2006;  Ivanova et~al.\ 2007). 
In particular, we find in agreement with previous work
(J04; S07) that the dependence of the expected number of LMXBs $\lambda_t$ 
on encounter rate, $\lambda_t \propto \Gamma^{0.85}$, is shallower than the 
naively expected linear behavior.

\acknowledgements

Support for program GO-10597 was provided through a grant from STScI
which is operated by AURA, Inc., under NASA contract NAS5-26555.
Support for this work was provided by NASA through {\it Chandra} award
GO7-8105X issued by CXO, which is operated by the SAO
for and on behalf of NASA under contract NAS8-03060. 
AJ and DM thank FONDAP 15010003.


\clearpage


\end{document}